\documentclass[12pt,a4paper]{article}
\pdfoutput=1

\usepackage[a4paper]{geometry}
\usepackage{amsmath}
\usepackage{amsfonts}
\usepackage{amssymb}
\usepackage[nosort]{cite}
\usepackage[backref=page]{hyperref}
\usepackage{graphicx}
\usepackage[margin=20pt,font=small,labelfont=bf]{caption}

\newcommand{\eq}[1]{\begin{equation}
                     \begin{split} #1 \end{split}
                     \end{equation}}
\newcommand{\ov}{\overline}
\newcommand{\op}{\hspace{1pt}}

\allowdisplaybreaks[2]
\numberwithin{equation}{section}
\begin{document}

\vspace*{\stretch{1}}

\begin{center}
{\LARGE
Mass spectrum of type IIB flux compactifications \\ --- \\
comments on AdS vacua and \\ conformal dimensions \\
}
\end{center}

\vspace{0.6cm}

\begin{center}
Erik Plauschinn
\end{center}

\vspace{0.6cm}

\begin{center} 
\textit{
Institute for Theoretical Physics, Utrecht University \\
Princetonplein 5, 3584CC Utrecht \\
The Netherlands \\
}
\end{center} 

\vspace{1cm}


\begin{abstract}
\noindent
In this note we study the mass spectrum of type IIB flux compactifications. 
We first give a general discussion of the mass matrix 
for F-term vacua in four-dimensional $\mathcal N=1$ supergravity 
theories and then specialize to type IIB Calabi-Yau orientifold compactifications
in the presence of geometric and non-geometric fluxes. 
F-term vacua in this setting are in general AdS$_4$ vacua for which 
we compute the conformal dimensions of operators dual to the scalar fields. 
For the mirror-dual of the DGKT construction we find that  one-loop corrections 
to the complex-structure moduli space lead to real-valued conformal dimensions
 ---
only when ignoring these corrections we recover the integer values
previously  reported in the literature.
For an example of a flux configurations more general than the DGKT mirror
we also obtain non-integer conformal dimensions.
Furthermore, we argue that stabilizing moduli in asymptotic regions of moduli space 
implies that at least one of the mass eigenvalue diverges. 
\end{abstract}

\vspace*{\stretch{1}}


\clearpage
\tableofcontents


\section{Introduction}

String theory is a theory of quantum gravity including gauge interactions.
Its supersymmetric version is consistent in ten  space-time dimensions
and
can be connected to four-di\-men\-sion\-al physics
through compactification.  
Well-understood compactification spaces are Calabi-Yau three-folds
whose resulting lower-di\-men\-sion\-al theories have a Minkowski vacuum. 
Deformations that preserve the Calabi-Yau condition 
correspond to massless scalar fields (moduli) in four dimensions, 
however, it is possible to deform the compact space away from being Calabi-Yau
by turning on fluxes. Such fluxes generate mass terms for the moduli 
and
typically lead to AdS$_4$ vacua.
For gravity theories in AdS spaces one can then apply the AdS/CFT 
dictionary to relate, for instance, the masses of scalar fields in 
AdS$_4$ to conformal dimensions of corresponding operators 
in a putative three-dimensional  CFT.
For a certain class of type II flux compactifications, 
more concretely for the construction by
DeWolfe, Giryavets, Kachru, and Taylor (DGKT)  in type IIA string theory \cite{DeWolfe:2005uu},
it was observed in \cite{Conlon:2021cjk,Apers:2022zjx,Apers:2022tfm,Quirant:2022fpn}
that the conformal dimensions of all scalar fields in the closed-string sector are
integer-valued. 
In \cite{Conlon:2021cjk,Apers:2022zjx} the analysis was done for toroidal compactifications, 
while in \cite{Apers:2022tfm,Quirant:2022fpn} the authors performed their computation 
for a general Calabi-Yau three-fold.

Obtaining integer conformal dimensions irrespective of the
compactification space  is a rather surprising observation which one would like to understand. 
In particular, one would like to know which features 
of the compactification are relevant for this result.
This question was addressed in \cite{Quirant:2022fpn}, where 
it was found that some non-supersymmetric type IIA vacua 
do not lead to integer-valued conformal dimensions. 
However, one can argue that such configurations are unstable \cite{Marchesano:2021ycx}
and therefore are not suitable for the applying the AdS/CFT correspondence \cite{Quirant:2022fpn}. 
In this note we approach the question of integer conformal dimensions
from the type IIB side. 
We study orientifold compactifications of type IIB string theory 
with geometric and non-geometric fluxes that lead to 
supersymmetric AdS$_4$ vacua. 
In regard to the DGKT construction in type IIA string theory \cite{DeWolfe:2005uu} 
we note the following differences:
\begin{enumerate}

\item On the type IIA side one typically considers the 
large-volume limit  without  corrections.
In type IIB string theory
this limit  corresponds to the large-complex-structure 
limit, for which  corrections  are  well-understood.
We can therefore compute conformal dimensions in a setting that includes  corrections
to the  moduli-space geometry.

\item For  DGKT flux compactifications in type IIA string theory the superpotential 
splits into the sum of two independent terms \cite{Grimm:2004uq}.
From a type IIB perspective this  is a rather special case that  corresponds 
to turning on only a specific component of geometric and non-geometric 
Neveu-Schwarz--Neveu-Schwarz fluxes. 
For a more general flux-choice the superpotential will not split in this way.

\end{enumerate}

In this work we investigate these two aspects. First, 
we consider the mirror-dual of the DGKT setting --- 
including perturbative corrections to the   complex-structure moduli space ---  
and determine  masses and conformal dimensions
analytically.
Second, we construct a flux configuration for which 
the superpotential does not split into two separate terms 
and determine the masses and conformal dimensions numerically. 
More concretely, 
\begin{itemize}

\item in section~\ref{sec_gen} we give a general discussion 
of the mass matrix for F-term vacua of four-dimensional $\mathcal N=1$ 
supergravity theories. In section~\ref{sec_iib} we then specialize to 
Calabi-Yau
orientifold  compactifications
of type IIB string theory with O3- and O7-planes in the 
presence of geometric Ramond-Ramond (R-R) and geometric and non-geometric Neveu-Schwarz--Neveu-Schwarz
(NS-NS) fluxes.

\item In section~\ref{sec_conf_dim} we determine masses and conformal dimensions 
for AdS$_4$ flux-vacua. For the  mirror-dual of DGKT
we find analytically that perturbative corrections to the prepotential lead to non-integer conformal dimensions. 
We also study an example with a flux choice  more general  than DGKT for which the superpotential does not split into two 
separate terms. Here we find numerically that the conformal dimensions are not integer.

\item Section~\ref{app_gen} is independent of our discussion of AdS vacua and conformal dimensions,
but uses many results from sections~\ref{sec_gen} and \ref{sec_iib}.
Here we study flux compactifications relevant for the KKLT and large-volume scenarios
\cite{Kachru:2003aw,Balasubramanian:2005zx},
where only the F-terms of the complex-structure moduli and the axio-dilaton 
are considered. 
We compute the trace of the canonically-normalized mass matrix 
and argue that if moduli are stabilized in an asymptotic regime of  moduli space, then 
at least one of the mass eigenvalues diverges.

\item In section~\ref{sec_conc} we summarize our findings and
in  appendix~\ref{app_conf} we give some technical details of the computations 
in the main text.

\end{itemize}


\section{Mass matrix and conformal dimensions}
\label{sec_gen}

Let us start with a  discussion of masses of  chiral multiplets 
in four-di\-men\-sio\-nal $\mathcal N=1$ 
supergravity theories. We consider minima of the scalar potential 
corresponding to vanishing F-terms and determine the general form of the 
mass matrix of the scalar fields. In the case of AdS$_4$ vacua we furthermore compute the conformal dimension of 
operators dual to the scalar fields.


\subsubsection*{F-term minima}

Let us consider a four-dimensional $\mathcal N=1$ supergravity theory
with $n$ complex scalar fields $\phi^M$ where $M=1,\ldots,n$. The  F-term scalar potential 
can be written as
\eq{
  \label{pot_f}
  V = e^K \left[  F_{M\vphantom{\ov N}} \op G^{M\ov N}\op \ov F_{\ov N} - 3 \op\lvert W\rvert^2\right],
}
where $K$ denotes the real K\"ahler potential,  $W$ denotes the holomorphic superpotential, and 
$G_{M\ov N} = \partial_{M\vphantom{\ov N}}\partial_{\ov N} K$ denotes the K\"ahler metric. 
The F-terms are given by $F_M=\partial_M W + K_M\op W$ with $K_M = \partial_M K$.
In this work we are interested in F-term minima of this potential given by 
\eq{
\label{min_f}
  F_M = 0\,.
}


\subsubsection*{Mass matrix}

The mass matrix for the complex scalar fields of this theory corresponds to  the second derivatives of the potential \eqref{pot_f}.
It can be arranged into the  form
\eq{
  \label{mass_001}
  m^2 = \left[ \begin{array}{cc} m^2_{M\ov N} & m^2_{MN} \\[4pt] m^2_{\ov M \ov N} & m^2_{\ov M N}
  \end{array}
  \right],
}
where the blocks in the first line are related to the ones appearing in the second line by complex conjugation.
For the former we find the following expressions at the minimum \eqref{min_f}
\eq{
  \label{mass_39856}
  &m^2_{M\ov N} = e^K \left[ \partial_{M\vphantom{\ov Q}} F_{P\vphantom{\ov Q}} \,G^{P\ov Q} \,\partial_{\ov N} \ov F_{\ov Q} -2\op 
   G_{M\ov N} \op \lvert W \rvert^2 \right] ,
  \\
  &m^2_{MN} = e^K \left[ - 2 \op\partial_M F_N \op \ov W \right] .
}
In order to obtain the canonically-normalized mass matrix we note that the K\"ahler metric 
$G_{M\ov N}$ is hermitian and positive definite. 
We can therefore write $G$ as the square of a positive-definite matrix $\Gamma$ and we define a matrix $Q$ as
\eq{
  \label{exp_384}
  G= \Gamma\op \Gamma^{\dagger}\,,
  \hspace{70pt}
  Q =    \Gamma{}^{-1} (\partial F) \op \Gamma{}^{-T} \,.
}
Here and in the following we suppress identity matrices $\delta_{M\ov N}$, $\delta^{M\ov N}$, and $\delta_M{}^N$.
The ca\-no\-nical\-ly-normalized mass matrix is obtained by multiplying \eqref{mass_001} with 
appropriate factors of $\Gamma$ 
from the left and the right and we find
\eq{
  \label{mass_can}
  m^2_{\rm can} =  
   e^K  \left[ \begin{array}{cc} Q  Q^{\dagger} -2 \op |W|^2 & - Q \op\ov W \\ -  Q^{\dagger}\op W & Q^{\dagger} Q -2\op |W|^2
  \end{array}
  \right].
}


\subsubsection*{Mass eigenvalues}

In order to determine the eigenvalues of \eqref{mass_can} we first perform a singular-value decomposition of $Q$ as
\eq{
 Q = U\op \Sigma\op V^{\dagger} \,, 
}
where $U$ and $V$ are unitary matrices and $\Sigma$ is a diagonal matrix that contains only real entries. 
We can then write the canonically-normalized mass matrix \eqref{mass_can} as
\eq{
m^2_{\rm can}  = e^K
\left[\begin{array}{cc} U & 0 \\ 0 & V \end{array}\right]
\left[\begin{array}{cc} \Sigma^2 -2\op |W|^2 & -\Sigma \, W \\ 
-\Sigma \, W & \Sigma^2 -2\op |W|^2  \end{array}\right]
\left[\begin{array}{cc} U^{\dagger} & 0 \\ 0 & V^{\dagger} \end{array}\right],
}
and the corresponding eigenvalue equation for the mass eigenvalues $\mathsf m^2$  is given by
\eq{
\label{det_010}
0 &= \det \op \Bigl[m^2_{\rm can} - \mathsf m^2\Bigr]\\
& = \det\op\Bigl[ e^{2K} \Bigl[ (\Sigma^2 -2\op \vert W\vert^2 - e^{-K}\mathsf m^2)^2
-\Sigma^2\op \vert W\vert^2 \Bigr] \Bigr] .
}
Denoting the entries of the diagonal matrix $\Sigma^2$ by $\sigma^2_{\alpha}$, we can solve
\eqref{det_010} as
\eq{
\label{exp_924794537}
  \mathsf m^2_{\alpha\pm} = e^K \Bigl[\op \sigma^2_{\alpha}  \pm \sigma^{\vphantom2}_{\alpha} \op \lvert W\rvert - 2 \op \lvert W\rvert ^2 \Bigr]\,.
}
We finally note that  $Q Q^{\dagger} = U\op  \Sigma^2 U^{\dagger}$
and $Q^{\dagger} Q = V\op \Sigma^2 V^{\dagger}$, and hence
$\sigma^2_{\alpha}$ are the real and positive eigenvalues of 
the hermitian matrices $Q^{\dagger} Q$ and $QQ^{\dagger}$.


\subsubsection*{AdS vacua and conformal dimensions}

If the superpotential $W$ is non-vanishing at the minimum, the 
F-term vacua of \eqref{pot_f} are AdS$_4$ vacua.
The corresponding AdS radius is defined as
\eq{
  R^2_{\rm AdS} = - \frac{3}{V\rvert_{\rm min}} = \frac{e^{-K}}{\lvert W \rvert^2 } \,,
}
and the mass eigenvalues \eqref{exp_924794537} can be brought into the form
\eq{
  \mathsf m^2_{\alpha\pm} = e^K \bigl[\sigma_{\alpha}  \pm  \tfrac{1}{2}\op \lvert W\rvert\bigr]^2  - \frac{9}{4}\op 
  \op\frac{1}{ R^2_{\rm AdS}}\,.
}
Hence, as expected,  the Breitenlohner-Freedman bound 
\raisebox{0pt}[0pt][0pt]{$\mathsf m^2_{\rm can}\geq -\frac{(d-1)^2}{4}\op R^{-2}_{\rm AdS}$} \cite{Breitenlohner:1982jf}
is satisfied for these  vacua. 
Furthermore, the masses of the scalar fields are related to the conformal dimensions
$\Delta$ of operators in a dual CFT as
$\Delta (\Delta-d) = \mathsf m^2_{\rm can} \op R^2_{\rm AdS}$.
For $d=3$ we then obtain
\eq{
  \Delta = \frac{1}{2} \left[ 3\pm \sqrt{9+ 4\op \mathsf m^2_{\rm can} \op R^2_{\rm AdS}} \right] ,
}
and using \eqref{exp_924794537} for the canonically-normalized masses we have 
$\Delta = 1 \pm \left\lvert {\sigma_{\alpha}}/{ W}\right\rvert$
and
$\Delta = 2 \mp \left\lvert {\sigma_{\alpha}}/{ W}\right\rvert$.
In order to satisfy the unitarity bound for all values of $\vert \sigma_{\alpha}/W\rvert$ 
one should choose the upper sign in the first expression and the lower sign 
in the second one, however, in principle the opposite sign choice is allowed as well. 
Here we make  the choice
\eq{
  \label{conf_dim}
  \Delta_{\alpha(1)} = 1 + \left\lvert \frac{\sigma_{\alpha}}{ W}\right\rvert , 
  \hspace{70pt}
  \Delta_{\alpha(2)} = 2 + \left\lvert \frac{\sigma_{\alpha}}{ W}\right\rvert,
}
for which  the conformal dimensions of the dual operators come in pairs that differ by one.


\section{Type IIB flux compactifications}
\label{sec_iib}

In this section we  consider four-dimensional $\mathcal N=1$
super\-gra\-vity theories that originate from type IIB flux compactifications on 
Calabi-Yau orientifolds. Our main result in this section 
is an expression for the matrix $Q$ appearing in the canonically-normalized
mass matrix \eqref{mass_can}.


\subsubsection*{Scalar fields}

Let us consider compactifications of type IIB string theory on 
Calabi-Yau three-folds $\mathcal X$, subject to an orientifold projection
leading to O3- and O7-planes \cite{Grimm:2004uq,Kachru:2004jr}.
This projection splits the cohomology of $\mathcal X$ into even and odd 
eigenspaces $H^{p,q}_{\pm}(\mathcal X)$ whose dimensions will be denoted by $h^{p,q}_{\pm}$. 
The resulting effective four-dimensional theory 
contains two classes of scalar fields:
first, there are
$h^{1,1}+1$ K\"ahler-sector moduli 
written as (we mostly follow the conventions of \cite{Plauschinn:2020ram})
\eq{
  \label{def_moduli}
  \mathsf T^A = (\op \tau, G^{\alpha}, T_a) \,, \hspace{50pt} 
  A=0,\ldots, h^{1,1}\,,
}
where $\tau$ denotes the axio-dilaton, $G^{\alpha}$ with $\alpha=1,\ldots, h^{1,1}_-$ 
are axionic moduli, and $T_a$ with $a=1, \ldots, h^{1,1}_+$ are the ordinary K\"ahler moduli
(see for instance \cite{Grimm:2004uq} for details). 
Our conventions are such that $\mbox{Re}\op \mathsf T^A$ are axionic degrees of freedom.
Second, there are $h^{2,1}_-$ complex-structure moduli $z^i$ with $i=1,\ldots, h^{2,1}_-$.
These are contained in the holomorphic three-form $\Omega$ of the Calabi-Yau three-fold.
Choosing a symplectic basis $\{\alpha_I,\beta^I\}\in H^3_-(\mathcal X)$ with $I=0,\ldots, h^{2,1}_-$, 
the holomorphic three-form can be expanded as
\eq{
  \label{hol_3}
  \Omega = X^I \alpha_I - \mathcal F_I\op \beta^I \,, \hspace{50pt}
  z^i = \frac{X^i}{X^0} \,,
}
where the periods $\mathcal F_I$ depend holomorphically on the complex-structure moduli $z^i$. 
We finally note that the effective theory  also contains $h^{2,1}_+$ vector fields
which may give rise to a D-term potential, however, here we assume that the D-term potential vanishes.


\subsubsection*{K\"ahler potential}

The dynamics of the scalar fields is determined by the K\"ahler potential. For the above setting it is given by
\eq{
  \label{lu_005}
  K  =  K_{\rm K} + K_{\rm cs} \,,
  \hspace{50pt}
  \arraycolsep2pt
  \begin{array}{lcl}
  K_{\rm K} &=& \displaystyle  -\log\left[\vphantom{\bigl(} - i\op (\tau-\ov \tau)\right]   -2 \log\left[ \mathcal{V} + \frac{\xi}{2} \right],  \\[10pt]
  K_{\rm cs} &=& \displaystyle  -\log\left[ +\op i \int_{\mathcal X}\Omega\wedge \ov \Omega \op\right],
  \end{array}
}
where 
the Einstein-frame volume of the Calabi-Yau three-fold is denoted by $\mathcal V$,
which depends implicitly on  
$\tau$, $G^{\alpha}$ and $T_a$, and we included 
$\alpha'$-corrections  encoded in
\raisebox{0pt}[0pt][0pt]{$\xi = -\frac{\zeta(3)\op\chi(\mathcal X)\op(\tau-\ov\tau)^{3/2}}{2\op(2\pi)^3\op(2i)^{3/2}}$} 
\cite{Becker:2002nn}. 
The K\"ahler potential for the K\"ahler-sector moduli --- including the $\alpha'$-corrections shown above --- 
has some special properties. With
$K_A =\partial_A K$, $G_{A\ov B} = \partial_{A\vphantom{\ov B}}\partial_{\ov B}K$,
and $K^A =G^{A\ov B}K_{\ov B}$ 
one finds
\eq{
  \label{lu_006}
 K^A = - (\mathsf T- \ov{\mathsf{T}})^A\,, 
 \hspace{50pt} K_{A\vphantom{\ov B}}\op G^{A\ov B} K_{\ov B} = 4\,.
}
Note, however, that the complex-structure sector does not satisfy 
similar relations in general. Only in certain limits one may find for instance $K_{i\vphantom{\ov j}} \op G^{i\ov j} K_{\ov j}=3$;
we come back to this point below.


\subsubsection*{Fluxes}

We furthermore consider fluxes along the compact space $\mathcal X$. They
generate a scalar potential in the four-dimensional theory which can be 
described by the superpotential \cite{Shelton:2005cf,Villadoro:2006ia,Shelton:2006fd}
\eq{
  \label{pot_w}
  W = \int_{\mathcal X} \Omega\wedge \left( F_3- \Xi_A \op \mathsf T^A \right) .
}
Here $F_3$ is the R-R three-form flux and $\Xi_A$ are
geometric and non-geometric NS-NS three-form fluxes. 
In particular, $\Xi_0$ is the ordinary $H_3$-flux, 
$\Xi_{\alpha}$ correspond to geometric $F$-fluxes, and $\Xi^a$ correspond to non-geometric $Q$-fluxes
 \cite{Dasgupta:1999ss,Kachru:2002sk,Hull:2004in} 
 (see \cite{Plauschinn:2018wbo} for a review). 
All fluxes are integer quantized.
The Bianchi identities for the NS-NS fluxes read 
\cite{Shelton:2005cf,Grana:2006hr,Plauschinn:2020ram}
\eq{
  \int_{\mathcal X} \Xi_A \wedge \Xi_B = 0 \,, 
}
while the Bianchi identities for the R-R fields contain  contributions of localized sources
such as D-branes and orientifold planes. 
The integrated versions of the Bianchi identities are known as tadpole-cancellation conditions. 
The fluxes contribute as \cite{Plauschinn:2020ram}
\eq{
  \label{tadpole_charges}
  N_A = \int_{\mathcal X} F_3 \wedge \Xi_A \,,
}
where $A=0$ corresponds to the D3-brane tadpole, $A = \alpha$  to
the D5-brane tadpole, and $A=a$ to the D7-brane tadpole.


\subsubsection*{Towards canonically-normalized-mass eigenvalues}

In order to compute the eigenvalues of the canonically-normalized mass matrix
shown in \eqref{exp_924794537}, we need to determine the eigenvalues 
$\sigma^2_{\alpha}$ of the matrices $Q Q^{\dagger}$ or
$Q^{\dagger} Q$.
To do so, we first define the matrix
\eq{
  \label{mat_qqq}
  \mathsf Q = G^{-1} \op \partial F \,,
}
evaluated at the minimum.
Since the matrix $\Gamma$ appearing in \eqref{exp_384} 
is invertible, we see that $\mathsf Q \ov{\mathsf Q}$ and $QQ^{\dagger}$ 
have the same eigenvalues. Let us therefore determine 
$\mathsf Q$ for the above setting:
we denote the K\"ahler-covariant derivative with respect to the complex-structure 
moduli $z^i$ by $D_i$. Its action on the $(3,0)$-form $\Omega$
leads to $(2,1)$-forms 
$\chi_i = D_i \op\Omega  \in H^{2,1}_-(\mathcal X)$
and its triple action on $\Omega$ leads to the Yukawa couplings $\kappa_{ijk}$
 \cite{Strominger:1990pd,Candelas:1990pi}.
More concretely, we have
(we follow the conventions of \cite{Candelas:1990pi})
\eq{
  \chi_i = \partial_{i} \Omega + K_i \op \Omega\,, 
  \hspace{50pt}
   \kappa_{ijk} = -\int_{\mathcal X} \Omega \wedge D_iD_jD_k \op\Omega\,.
}
We also note that the $(1,2)$-components of the real three-forms $\Xi_A$ are given by
\eq{
  \label{rel_3846}
  \Xi_A^{\ov i} = -i\op e^{K_{\rm cs}}\op G^{\ov i j}\int_{\mathcal X} \chi_j \wedge \Xi_A \,,
}
which are related to the $(2,1)$-components $\Xi_{\ov A}^{i}$ by complex conjugation. 
Using then the F-term conditions $F_A=0$ and $F_i=0$ as well as
special-geometry relations of the complex-structure moduli space  \cite{Strominger:1990pd,Candelas:1990pi},
we can determine 
the matrix $\mathsf Q$ shown in \eqref{mat_qqq} as follows (details of this computation are shown in appendix~\ref{app_details})
\eq{
  \label{res_q_001}
  \arraycolsep2pt
  \renewcommand{\arraystretch}{1.4}
   \begin{array}{lcl}
  \displaystyle \mathsf Q^{\ov A}{}_B &=& \displaystyle \bigl[ - \delta^{\ov A}{}_B - K^{\ov A}K_B\bigr] \op W\,,
  \\
  \displaystyle \mathsf Q^{\ov A}{}_j &=& \displaystyle -i \op e^{-K_{\rm cs}} \op G^{\ov A B}  \, \Xi^{\ov i}_B\op G_{\ov i j}\,,
  \\
  \displaystyle \mathsf Q^{\ov i}{}_B &=& \displaystyle -i \op e^{-K_{\rm cs}} \, \Xi^{\ov i}_B \,,
  \\
  \displaystyle \mathsf Q^{\ov i}{}_j &=& \displaystyle  G^{\ov i m}\kappa_{jmn}\, \Xi^{ n}_{\ov B} \op K^{\ov B} \,.
  \end{array} 
}
Note that these expressions are valid for any point in complex-structure moduli space.
In principle one can use them to either compute the singular-value decomposition 
of $\mathsf Q$ and determine the singular values $\sigma_{\alpha}$ or to 
compute the eigenvalues $\sigma^2_{\alpha}$ of $\mathsf Q\ov{\mathsf Q}$.
However, we were not able to obtain analytic expressions for the eigenvalues for general flux configurations.


\subsubsection*{Remarks}

We close this section with two remarks.
First, using \eqref{lu_006} and the F-term condition of the K\"ahler-sector moduli,  
from $K^A F_A =0$ we determine
\eq{
  \label{w_pot_min_24}
  W \rvert_{\rm min} = -\frac{i}{2} \int_{\mathcal X} \Omega\wedge \Xi_A \op (\mbox{Im}\op \mathsf T^A) \,.
}
Second, with the help of  the F-term conditions  we can express the tadpole charges 
\eqref{tadpole_charges} at the minimum as 
\eq{
  N_A = \left[ \int_{\mathcal X} \Xi_{A\vphantom{\ov B}} \wedge \star\op \Xi_{\ov B} + 12 \op G_{A\ov B}\, e^{K_{\rm cs}} \lvert W\rvert^2 \right] 
  (\mbox{Im}\op \mathsf T^B) \,.
}
At the minimum the matrix in parenthesis is semi-positive definite, which 
implies in particular that $ N_A \op (\mbox{Im}\op \mathsf T^A)  \geq 0 $.


\section{Conformal dimensions}
\label{sec_conf_dim}

Since it is difficult to determine the eigenvalues of the mass matrix for general flux choices 
analytically, in this section
we consider two specific settings. First, we study the mirror-dual of 
the  type IIA DGKT construction
\cite{DeWolfe:2005uu}. 
Second, we analyze numerically an example with  geometric and 
non-geometric NS-NS fluxes for $h^{1,1}=h^{2,1}_-=1$.


\subsection{The type IIA mirror}
\label{sec_iia_mirror}

We start with a setting that is mirror-dual to the type IIA DGKT construction \cite{DeWolfe:2005uu}.
This configuration is special since 
the superpotential splits into a sum of two terms that only depend on the complex-structure and 
only on the K\"ahler-sector moduli, respectively.


\subsubsection*{Fluxes}

The setting that we consider on the type IIB side is characterized by the 
following choice of NS-NS fluxes
\eq{
  \label{iia_001}
  \Xi_A = - \left( \Xi_{ A}\right)_0 \beta^0 \,,
}
where we expanded $\Xi_A$ into the symplectic basis $\{\alpha_I,\beta^I\} \in H^3_-(\mathcal X)$. 
On the type IIA side $ \left( \Xi_{ A}\right)_0$ are the components of the $H_3$-flux, 
and we remark that the R-R three-form flux $F_3$ is not restricted besides the 
tadpole-cancellation condition. 
From the expansion of the holomorphic three-form shown in \eqref{hol_3} we see
that the superpotential \eqref{pot_w} indeed splits into two terms depending 
only on the complex-structure and only on the K\"ahler-sector moduli
\eq{
  W = \int_{\mathcal X} \Omega\wedge F_3 + \Bigl[ X^0  \left( \Xi_{ A}\right)_0 \mathsf T^A \Bigr] \,.
}
Furthermore, from \eqref{w_pot_min_24} we see that in this case  $W/X^0$ at the minimum is purely 
imaginary which matches the discussion in \cite{DeWolfe:2005uu}.
Using then $\int_{\mathcal X}\partial_i \Omega \wedge \Xi_A = 0$ and 
the F-term condition $F_A=0$, we find for the $(1,2)$-components of $\Xi_A$ 
\eq{
  \label{iia_002}
  \Xi_A^{\ov i} = -i \op e^{K_{\rm cs}} \op K_A \op K^{\ov i} \op W \,.
}


\subsubsection*{Large-complex-structure limit}

For the type IIA setting one  considers the large-volume limit in which  corrections 
are typically neglected. On the type IIB side this limit 
corresponds to the large-complex-structure limit, where 
subleading corrections are however well-understood. 
In particular,  the holomorphic three-form is determined by 
the following prepotential at the perturbative level
\eq{
\label{ppot}
\mathcal F = - \frac{1}{3!} \op\frac{\kappa_{ijk} X^i X^j X^k}{X^0} + 
\frac{1}{2!} \op a_{ij} X^i X^j + b_i X^i X^0 + 
\frac{1}{2!} \op c \op
(X^0)^2  \,,
}
where $a_{ij}$ and $b_i$ are real while $c$ is purely imaginary. 
In this work we ignore instanton corrections to the prepotential.
The periods $\mathcal F_I$ appearing in 
\eqref{hol_3} are given by $\mathcal F_I = \partial_{X^I} \mathcal F$,
from which one  can determine
the K\"ahler potential, the first derivatives $K_i$, and the K\"ahler metric
$G_{i\ov j}$. 
Let us  define
\eq{
  z^i = u^i + i\op v^i\,, \hspace{70pt}
  \gamma = \frac{3\,\mbox{Im}\op c}{\kappa_{ijk} v^i v^j v^k}\ll 1\,,
}
and note that in the large-complex-structure limit we have $\gamma\to 0$.
We then compute the following expressions
\begin{subequations}
 \label{iia_003}
\begin{align}
  &K^i = -i \,\frac{2-\gamma}{1+\gamma} \op v^i\,,
  \\
  &
 \label{iia_003b}
  K_i \op G^{i\ov j}  K_{\ov j} = \frac{3}{1+\gamma} \,,
  \\
  &
    \kappa^{\ov i}{}_{jk} K^k = \frac{i}{2} \op \frac{X{}^0}{\ov X{}^0}\op  e^{-K_{\rm cs}} \op  \frac{2-\gamma}{1+  \gamma} 
  \left[ \delta^{\ov i}{}_j + K^{\ov i}K_j \right] .
\end{align}
\end{subequations}


\subsubsection*{The eigenvalues of $QQ^{\dagger}$}

With the help of \eqref{iia_002} and \eqref{iia_003} we  determine the matrix 
$\mathsf Q$ and subsequently $\mathsf Q\ov{\mathsf Q}$. 
Noting that \eqref{w_pot_min_24} together with \eqref{iia_001} implies
that $W/X^0$ is purely imaginary at the minimum, we obtain the following  four sub-blocks 
\eq{
  \label{qq_iia}
  \arraycolsep2pt
  \renewcommand{\arraystretch}{2.5}
   \begin{array}{lcl}
  \displaystyle (\mathsf Q\ov{\mathsf Q})^{\ov A}{}_{\ov B} &=& \displaystyle \lvert W\rvert^2 \left[ \delta^{\ov A}{}_{\ov B} +\frac{5+2\op \gamma}{1+ \gamma} \op K^{\ov A}K_{\ov B}\right] ,
  \\
  \displaystyle (\mathsf Q\ov{\mathsf Q})^{\ov A}{}_{\ov j} &=& \displaystyle \lvert W\rvert^2 K^{\ov A}K_{\ov j} \left[ 3- 2\left( \frac{2-\gamma}{1+\gamma}\right)^2\right] ,
  \\
  \displaystyle (\mathsf Q\ov{\mathsf Q})^{\ov i}{}_{\ov B} &=& \displaystyle \lvert W\rvert^2 K^{\ov i}K_{\ov B} \left[ 3- 2\left( \frac{2-\gamma}{1+\gamma}\right)^2\right] ,
  \\
  \displaystyle (\mathsf Q\ov{\mathsf Q})^{\ov i}{}_{\ov j} &=& \displaystyle \lvert W\rvert^2\left[ 4\left( \frac{2-\gamma}{1+\gamma}\right)^2 \delta^{\ov i}{}_{\ov j}
  +\left( 4+4\left(\frac{2-\gamma}{1+\gamma}\right)^2  \frac{1-2\gamma}{1+\gamma} \right) K^{\ov i}K_{\ov j} 
  \right] .
  \end{array} 
}
These expressions are in line with corresponding type IIA formulas found in 
\cite{
Herraez:2018vae,
Marchesano:2019hfb,
Apers:2022tfm
}.
In order to determine the eigenvalues $\sigma^2_{\alpha}$ of $\mathsf Q\ov{\mathsf Q}$, 
we first note that $K_A$ is a $(h^{1,1}+1)$-dimensional vector and that there 
are $h^{1,1}$ vectors $L_{\ov A}$ perpendicular to $K_A$ with respect to the K\"ahler metric, i.e.~they satisfy $K_A \op L^A = 0$.
Similarly, $K_i$ is a $h^{2,1}_-$-dimensional vector and there are $h^{2,1}_--1$ 
vectors $L_{\ov i}$ that satisfy $K_i \op L^i = 0$.
Using then \eqref{lu_006} and \eqref{iia_003b},
we can compute the  eigenvalues and eigenvectors of 
the matrix $\mathsf Q\ov{\mathsf Q}$. They are  summarized in table~\ref{ev_001}.
\begin{table}[p]
\begin{equation*}
  \arraycolsep10pt
  \renewcommand{\arraystretch}{1.5}
  \begin{array}{l@{\hspace{2pt}}c@{\hspace{2pt}}r@{\hspace{2pt}}c@{\hspace{2pt}}l || c || c}
  \multicolumn{5}{c ||}{\mbox{eigenvalues $\sigma^2_{\alpha}/\lvert W\rvert^2$} }& \mbox{eigenvectors} & \mbox{multiplicity}
  \\
  \hline\hline
    \multicolumn{5}{c ||}{1} & ( L^{\ov A}, 0)^T &h^{1,1} 
  \\
    \multicolumn{5}{c ||}{\left( \frac{4-2\gamma}{1+\gamma}\right)^2}  & (0,L^{\ov i})^T & h^{2,1}_- - 1
  \\
  16 &+&\frac{96}{13}\op\gamma&+&\mathcal O(\gamma^2) & ( K^{\ov A}, \eta_{(1)} \op K^{\ov i})^T& 1
  \\
  81 & -&\frac{5400}{13}\op \gamma &+&\mathcal O(\gamma^2) & ( K^{\ov A}, \eta_{(2)} \op K^{\ov i})^T & 1
  \end{array}
\end{equation*}
\caption{Eigenvectors and eigenvalues of the matrix $\mathsf Q\ov{ \mathsf Q}$ shown in 
 \eqref{qq_iia}. 
The vectors $L^{\ov A}$ and $L^{\ov i}$ satisfy $K_{\ov A}L^{\ov A}=0$ and 
$K_{\ov i}L^{\ov i}=0$, and the parameters $\eta_{(1)}$ and $\eta_{(2)}$ take the form
 $\eta_{(1)} = \frac{1}{3} + \frac{25}{39}\op\gamma+\mathcal O(\gamma^2)$ and
 $\eta_{(2)} = -4 + \frac{48}{13}\op\gamma+\mathcal O(\gamma^2)$.
 The precise expressions for the eigenvalues and eigenvectors in the last two lines 
 are shown in appendix~\ref{app_ev_details}. 
  \label{ev_001}
}
\end{table}


\subsubsection*{Masses and conformal dimensions}

Using the results shown in table~\ref{ev_001}, we can now determine the 
canonically-nor\-ma\-lized masses 
and corresponding conformal dimensions using 
\eqref{exp_924794537} 
and \eqref{conf_dim}.
The resulting expressions are summarized in table~\ref{md_001}, 
where we assumed that $W\neq 0$ at the minimum.
\begin{table}[p]
\begin{equation*}
  \arraycolsep10pt
  \renewcommand{\arraystretch}{1.5}
  \begin{array}{l@{\hspace{2pt}}c@{\hspace{2pt}}r@{\hspace{2pt}}c@{\hspace{2pt}}l || r@{\hspace{2pt}}c@{\hspace{2pt}}r@{\hspace{2pt}}c@{\hspace{2pt}}l || c}
   \multicolumn{5}{c ||}{\mbox{mass $\mathsf m^2 /R^{-2}_{\rm AdS}$}} & \multicolumn{5}{c ||}{\mbox{conformal dimension $\Delta$}} & \mbox{multiplicity}
  \\
  \hline\hline
   \multicolumn{5}{c ||}{0} & \multicolumn{5}{c ||}{2} & h^{1,1}  \\
   \multicolumn{5}{c ||}{-2}  & \multicolumn{5}{c ||}{3} & h^{1,1} \\
  \hline
    \multicolumn{5}{c ||}{\frac{18(1-\gamma)}{(1+\gamma)^2}} & \multicolumn{5}{c ||}{\frac{5-\gamma}{1+\gamma}} & h^{2,1}_--1  \\
   \multicolumn{5}{c ||}{\frac{2(5-\gamma)(1-2\gamma)}{(1+\gamma)^2}} & \multicolumn{5}{c ||}{\frac{6}{1+\gamma} }& h^{2,1}-1 \\
  \hline
   18&+& \frac{108}{13}\op \gamma &+& \mathcal O(\gamma^2)& 5 &+& \frac{12}{13}\op\gamma&+&\mathcal O(\gamma^2) & 1  \\
   10&+& \frac{84}{13}\op \gamma &+& \mathcal O(\gamma^2)&  6 &+& \frac{12}{13}\op\gamma&+&\mathcal O(\gamma^2) & 1 \\
  \hline
   88&-& \frac{5700}{13}\op \gamma &+& \mathcal O(\gamma^2)& \hspace{13pt} 10 &-& \frac{300}{13}\op\gamma&+&\mathcal O(\gamma^2)& 1  \\
   70&-& \frac{5100}{13}\op \gamma &+& \mathcal O(\gamma^2)&  11 &-& \frac{300}{13}\op\gamma&+&\mathcal O(\gamma^2)& 1 \\
  \end{array}  
\end{equation*}
\caption{Masses and conformal dimensions  \label{md_001}
for the mirror-dual of the type IIA DGKT setting. 
In the strict large-complex-structure limit $\gamma=0$ and one obtains integer-valued 
conformal dimensions.
}
\end{table}
Our main observation is that when taking into account 
the corrections to the large-complex-structure limit 
encoded in $\gamma$, the conformal dimensions of the 
dual operators are not integer-valued. Only in 
the strict large-complex-structure limit $\gamma=0$ 
we reproduce the integer conformal dimensions 
found in\cite{Conlon:2021cjk,Apers:2022zjx,Apers:2022tfm}.


\subsection{General fluxes for $h^{1,1}=h^{2,1}_-=1$}
\label{sec_numerics}

In this section we analyze a compactification with a minimal set of moduli but with a 
more general choice of 
fluxes. We stabilize  moduli in an AdS$_4$ vacuum at large complex structure, large volume,  
and weak coupling without taking into account perturbative corrections to the prepotential.
We then study the masses and 
corresponding conformal dimensions numerically.


\subsubsection*{Setting}

Let us consider a setting with two K\"ahler-sector moduli and one complex-structure modulus.
This  corresponds to  a compactification manifold and orientifold projection with  Hodge numbers
\eq{
  h^{1,1} = h^{1,1}_+ = 1\,, \hspace{50pt} h^{2,1}_-=1\,.
}
We furthermore assume that the complex-structure modulus is 
stabilized at large complex structure. Ignoring the 
perturbative corrections and choosing $\kappa_{111}=1$ for simplicity, the corresponding  prepotential \eqref{ppot} simplifies to
\eq{
\label{ppot_simp}
\mathcal F = - \frac{1}{3!} \op\frac{\left( X^1\right)^3}{X^0} \,.
}
In the K\"ahler sector we ignore $\alpha'$-corrections and we express the Einstein-frame volume
of the Calabi-Yau manifold in  terms of the K\"ahler modulus $T_1$ as
\eq{
  \mathcal V = \frac{1}{6} \left[ -i\op(T_1 - \ov T_1) \right]^{3/2} \,.
}


\subsubsection*{Fluxes}

Turning to the fluxes, we expand the R-R and NS-NS three-form fluxes in the symplectic basis 
$\{\alpha_I,\beta^I\}$ as $F_3 = f^I\op\alpha_I - f_I\op \beta^I$ and 
$\Xi_A = (\Xi_A)^I\op\alpha_I - (\Xi_A)_I\op\beta^I$.
Arranging the components $f^I$, $f_I$ and
$(\Xi_A)^I$, $(\Xi_A)_I$ into a vector and matrix, respectively, we make the following 
choice
\eq{
  F_3 = \left( \begin{array}{c} 60\\0\\0\\ -2\mathsf a^2(9+2\mathsf b)\end{array} \right),
  \hspace{40pt}
  \Xi_A = \left( \begin{array}{cc}
  0 & 0 \\
  0 & 10\mathsf b \\
  -2\mathsf a^2(2+\mathsf b) & -\mathsf a^2(12+\mathsf b) \\
  0 & 0 
  \end{array}\right),
}
where $\mathsf a^2\in\mathbb Z$ and $\mathsf b \in\mathbb Z$.
The corresponding tadpole charges \eqref{tadpole_charges} are determined as
\eq{
  N_A = \Bigl( 120\op \mathsf a^2(2+\mathsf b)\,,\; 40\op \mathsf a^2(18-3\mathsf b-\mathsf b^2) \Bigr)\,.
}  
Note that $\mathsf b=0$ corresponds to a setting that is mirror dual to a type IIA DGKT 
construction. In the absence of corrections to the prepotential, as we 
are considering here, for $\mathsf b=0$ we therefore expect integer conformal dimensions.


\subsubsection*{Moduli stabilization}

For the above K\"ahler potential, superpotential, and choice of fluxes we can now solve
the F-term conditions. These fix the moduli as
\eq{
  z^1 = i\op \mathsf a \,, \hspace{50pt}
  \mathsf T^0 = \tau = i\op \mathsf a \,, \hspace{50pt}
  \mathsf T^1 = T_1 = i\op \mathsf a \,,
}
and  the AdS radius takes the value
\eq{
\frac{1}{R^2_{\rm AdS}} = \frac{27(2+\mathsf b)^2}{\mathsf a} \,.
}
Let us discuss two particular choices for the parameter $\mathsf b$.
As mentioned above, for the mirror-dual of the DGKT setting we expect integer conformal dimensions.
And indeed, for $\mathsf b=0$ the eigenvalues of $QQ^{\dagger}$ 
are $\sigma^2_{\alpha}/\lvert W\rvert^2=( 1,16,81)$
which lead to the canonically-normalized masses and conformal dimensions
\eq{
  \arraycolsep1.5pt
  \begin{array}{rcl @{\hspace{4pt}}r r r r r r r r r r r @{\hspace{4pt}} l}
  \displaystyle \left.\frac{\mathsf m^2}{R^{-2}_{\rm Ads}} \right\vert_{\mathsf b=0}& = & 
  \displaystyle \bigl(  & 0 & , & -2 &,& 18 &,& 10 &,& 88 &,& 70 & \displaystyle \bigr)\,,
  \\[20pt]
  \left. \Delta \right\vert_{\mathsf b=0} & = & 
  \displaystyle \bigl(  & 2 & , & 3 &,& 5 &,& 6 &,& 10 &,& 11 & \displaystyle \bigr)\,.
  \end{array}
}
On the other hand, the choice $\mathsf b=1$ corresponds to a more general setting that is different
from the DGKT mirror. Here we obtain
$\sigma^2_{\alpha}/\lvert W\rvert^2=(0.91 , 13.45 , 42.16 )$
which leads to
\eq{
  \arraycolsep1.5pt
  \begin{array}{rcl @{\hspace{4pt}}r r r r r r r r r r r @{\hspace{4pt}} l}
  \displaystyle \left.\frac{\mathsf m^2}{R^{-2}_{\rm Ads}} \right\vert_{\mathsf b=1} & = & 
  \displaystyle \bigl(  & -0.14 & , & -2.05 &,& 15.12 &,& 7.78 &,& 46.66 &,& 33.67 & \displaystyle \bigr)\,,
  \\[20pt]
  \left. \Delta \right\vert_{\mathsf b=1} & = & 
  \displaystyle \bigl(  & 1.91 & , & 2.91 &,& 14.45 &,& 15.45 &,& 43.16 &,& 44.16 & \displaystyle \bigr)\,.
  \end{array}
}
In particular, for a choice of  fluxes that is slightly more general than the mirror-dual of the 
type IIA DGKT setting, the conformal dimensions of the operators dual to the scalar fields are not integer. 
A similar observation was made in \cite{Apers:2022zjx}.


\section{Asymptotic regions}
\label{app_gen}

Our discussion in this section is independent from our analysis of  AdS vacua and conformal dimensions, 
but it utilizes many results from sections~\ref{sec_gen} and \ref{sec_iib}.
We study the mass matrix for 
type IIB flux compactifications that 
are relevant for the KKLT and Large-Volume scenarios 
\cite{Kachru:2003aw,Balasubramanian:2005zx}.
Using the trace of this matrix, we argue that 
stabilizing moduli in an asymptotic 
region of moduli space implies that at least one mass eigenvalue
diverges. 
This computation was part of the master thesis \cite{Draijer:2022}
and has been verified numerically (in a slightly different setting) in the master thesis \cite{Rottier:2022}.


\subsubsection*{Scalar potential}

We consider type IIB flux compactifications with R-R and NS-NS three-form fluxes
$F_3$ and $H_3$. In this case only the dilaton $\tau$ and the complex-structure moduli $z^i$
appear in the superpotential and hence $W$ is independent of the remaining K\"ahler-sector 
moduli.
When ignoring the $\alpha'$-corrections to the K\"ahler potential $K_{\rm K}$ 
shown in \eqref{lu_005}, the K\"ahler-sector moduli (without the axio-dilaton) satisfy 
the no-scale condition
\eq{
  K_{A\vphantom{\ov B}}\op G^{A\ov B} K_{\ov B}  = 3 
  \hspace{50pt}\mbox{for} \hspace{10pt} A,B=1,\ldots, h^{1,1}\,.
}
In this case the scalar F-term potential can be brought into the 
form
\eq{
  V = e^K \, F_{M\vphantom{\ov N}} \op G^{M\ov N}\op \ov F_{\ov N} \,,
}
where $M,N$ label the axio-dilaton $\tau$ and the complex-structure moduli $z^i$
but not the remaining K\"ahler-sector moduli $G^{\alpha}$ and $T_a$.
In the following we are interested in F-term minima 
given by
\eq{
  F_M = 0\,,
}
but we ignore the F-terms corresponding to the moduli $G^{\alpha}$ and $T_a$.
In the KKLT and large-volume scenarios these are stabilized in a second step using non-perturbative effects.


\subsubsection*{Mass matrix}

Following a discussion similar to the one in section~\ref{sec_gen}, we
find that the canonically-normalized mass matrix can be expressed as
\eq{
  \label{mass_can_2}
  m^2_{\rm can} =  
   e^K  \left[ \begin{array}{cc} Q  Q^{\dagger} + \op |W|^2 & 2\op Q \op\ov W \\ 2\op  Q^{\dagger}\op W & Q^{\dagger} Q +\op |W|^2
  \end{array}
  \right].
}
Denoting the eigenvalues of $QQ^{\dagger}$ again by $\sigma_{\alpha}^2$, we determine the 
eigenvalues of \eqref{mass_can_2} as
\eq{
 \mathsf m^2_{\alpha\pm} = e^K \bigl( \sigma_{\alpha}  \pm  \lvert W\rvert \bigr)^2\,.
}


\subsubsection*{The matrix $\mathsf Q\ov{\mathsf Q}$}

We  note that the eigenvalues of $QQ^{\dagger}$ are the same as the eigenvalues 
of $\mathsf Q\ov{\mathsf Q}$, where $\mathsf Q$ was defined in 
\eqref{mat_qqq}.
From the superpotential
\eq{
W = \int_{\mathcal X} \Omega\wedge \left( F_3- H_3 \op \tau \right) ,
}
we then compute
\eq{
  \mathsf Q^{\ov\tau}{}_{\tau}  &= 0\,, \\
  \mathsf Q^{\ov i}{}_{\tau} &= -i \op e^{-K_{\rm cs}} \op h^{\ov i} \,, \\
  \mathsf Q^{\ov \tau}{}_{j} &= +i \op(\tau-\ov \tau)^2\op  e^{-K_{\rm cs}} \op g_{j\ov i}\op h^{\ov i}  \,,\\
  \mathsf Q^{\ov i}{}_{j} & =(\tau-\ov\tau)\op  \kappa^{\ov i}{}_{jk} h^k \,,
}
where $h^{\ov i}$ with $i=1,\ldots, h^{2,1}_-$ are the $(1,2)$-components of $H_3$
(c.f.~equation \eqref{rel_3846}).
Denoting by $R_{i\ov j m\ov n}$ the Riemann tensor of the complex-structure moduli-space metric 
(we follow the conventions of \cite{Candelas:1990pi})
and by $K_\tau = -\log[-i(\tau-\ov\tau)]$ the K\"ahler potential of the axio-dilaton, we determine
\eq{
  \label{rel_1000}
  \left[  \mathsf Q\ov{\mathsf Q}\right]^{\ov \tau}{}_{\ov \tau} &= e^{-2(K_{\tau} + K_{\rm cs})} \op h^i g_{i\ov j} \op h^{\ov j} \,,
  \\
  \left[  \mathsf Q\ov{\mathsf Q}\right]^{\ov \tau}{}_{\ov j} &=e^{-K_{\tau} - K_{\rm cs}} (\tau-\ov\tau)^2 \op \kappa_{\ov j\ov m\ov n}
  h^{\ov m} h^{\ov n}\,,
  \\
  \left[  \mathsf Q\ov{\mathsf Q}\right]^{\ov i}{}_{\ov \tau} &=-e^{-K_{\tau} - K_{\rm cs}}  \op \kappa^{\ov i}{}_{ m n}
  h^{ m} h^{ n}\,,
  \\
  \left[  \mathsf Q\ov{\mathsf Q}\right]^{\ov i}{}_{\ov j} &= e^{-2(K_{\tau} + K_{\rm cs})} \left( 
  -R^{\ov i}{}_{\ov j m\ov n} h^m h^{\ov n} + \delta^{\ov i}{}_{\ov j} \op h^m g_{m\ov n} h^{\ov n}
  + 2 h^{\ov i}\op h_{\ov j}
  \right).
}


\subsubsection*{Trace of the canonically-normalized mass matrix}

The trace of the canonically-normalized mass matrix \eqref{mass_can_2} can be computed 
using \eqref{rel_1000}
as
\eq{
  \label{tr_01}
  \mbox{tr}\, m^2_{\rm can} = \frac{2}{\mathcal V^2} \Bigl[ 
   e^{-K_{\rm cs} - K_{\tau}} h^{m}  H_{m\ov n}  h^{\ov n}
  +e^{+K_{\rm cs} + K_{\tau}} (h^{2,1}+1)\lvert W\rvert ^2
  \Bigr]\,,
}
where all expressions are evaluated at the minimum. 
The Hodge metric 
$H_{i\ov j}$  
is defined in terms of the Ricci tensor 
$R_{i\ov j}$
on the complex-structure moduli space and  satisfies (see e.g.~\cite{Douglas:2006zj})
\eq{
  \label{tr_02}
  H_{i\ov j} = R_{i\ov j} + (h^{2,1}_-+3)\op g_{i\ov j} \,, \hspace{60pt}
  H_{i\ov j} \geq 2\op g_{i\ov j} \,.
}
We furthermore note that with the help of the F-term conditions the  flux number appearing in the D3-brane 
tadpole-cancellation condition can be written as
\eq{
N_{\rm flux}  = e^{-K_{\rm cs} - K_{\tau}} h^m g_{m\ov n} h^{\ov n} +  e^{+K_{\rm cs} + K_{\tau}}\lvert W\rvert^2 
\,,
}
and using the second relation in \eqref{tr_02} we find from 
\eqref{tr_01} the bound
\eq{
  \mbox{tr}\, m^2_{\rm can} \geq \frac{2}{\mathcal V^2} \Bigl[ 
   2\op N_{\rm flux}
  +e^{K_{\rm cs} + K_{\tau}} (h^{2,1}-1)\lvert W\rvert ^2
  \Bigr] .
}
Requiring $h^{2,1}_-\geq 1$ the bound above implies 
\eq{
  \label{trace_007}
  \mbox{tr}\, m^2_{\rm can} \geq \frac{4}{\mathcal V^2} \op N_{\rm flux} 
  \hspace{40pt}\Rightarrow  \hspace{40pt}
  \mathsf m^2_{\rm max} \geq  \frac{2}{\mathcal V^2} \op \frac{N_{\rm flux} }{h^{2,1}_-+1} \,,
}
where $\mathsf m^2_{\rm max}$ is the largest eigenvalue of the canonically-normalized mass matrix. 
(For $h^{2,1}_-=0$ we obtain $\mbox{tr}\, m^2_{\rm can}=2\op N_{\rm flux}/\mathcal V^2$.)
For the expression on the right-hand side in \eqref{trace_007} 
we used that $\mbox{tr}\, m^2_{\rm can}$ is the sum of all mass eigenvalues
and that $\mbox{tr}\, m^2_{\rm can}/2(h^{2,1}_-+1)$ is the average mass eigenvalue.
Note that in \cite{Conlon:2006gv} a similar relation for the average mass eigenvalue has been estimated, 
whereas here we give a precise derivation.


\subsubsection*{Moduli stabilization in asymptotic regions}

In the paper \cite{Plauschinn:2021hkp} we argued that when 
stabilizing moduli in the large-complex-structure limit by fluxes, the flux number 
$N_{\rm flux}$ is expected to diverge. In \cite{Grana:2022dfw}  this argument has been 
extended to arbitrary boundary limits using asymptotic Hodge theory. 
If this expectation is true, then
\eqref{trace_007} implies that  stabilizing moduli 
in an asymptotic region of moduli space means that at least one mass-eigenvalue 
will diverge --- provided that the overall volume $\mathcal V$ remains the same
\eq{
  N_{\rm flux} \xrightarrow{\hspace{4pt}\rm asymptotic~region\hspace{4pt}}\infty
  \hspace{40pt}\Rightarrow\hspace{40pt}
  \mathsf m^2_{\rm max} \xrightarrow{\hspace{4pt}\rm asymptotic~region\hspace{4pt}}\infty\,.
}
In a consistent string-theory compactification the flux number $N_{\rm flux}$ is bounded by the 
tadpole-cancellation condition, however, 
it has been argued that this bound can be rather large \cite{Crino:2022zjk}. Therefore, 
also $\mathsf m^2_{\rm max}$ can be large. 
In order to have a separation of scales between the moduli masses and the Kaluza-Klein masses
$\mathsf m^2_{\rm max}\ll m^2_{\rm KK}$, 
one  has to ensure a sufficiently large volume $\mathcal V$ and sufficiently small string coupling.
In particular, we have to require
\eq{
  \mathsf m^2_{\rm max} \ll m^2_{\rm KK}
  \hspace{40pt}\Rightarrow\hspace{40pt}
  \frac{N_{\rm flux}}{h^{2,1}_-+1} \ll 2\pi^2\, \mbox{Im}(\tau)\op \mathcal V^{2/3} \,.
}
For the large-volume scenario this is only a mild restriction, but 
for KKLT it may become relevant.


\section{Summary and conclusions}
\label{sec_conc}

In this note we  studied the mass spectrum of flux compactifications of type IIB
string theory. Let us  summarize and discuss our findings.


\subsubsection*{AdS vacua of type IIB flux compactifications}

In section~\ref{sec_gen} we considered F-term vacua of general four-dimensional
$\mathcal N=1$ supergravity theories and determined 
in equation \eqref{exp_924794537}
the eigenvalues $\mathsf m^2_{\alpha\pm}$ of the
canonically-normalized mass matrix for the scalar fields.
In general these vacua are AdS$_4$ vacua, for which we 
compute the conformal dimensions 
of  dual operators in a putative three-dimensional CFT. They 
are shown in \eqref{conf_dim}.
We find that both quantities are determined in a simple way 
by the value of the superpotential at the minimum and by
the singular values $\sigma_{\alpha}$ of the matrix
\eq{
  \mathsf Q^{\ov M}{}_{N} = G^{\ov M P}\partial_{P} F_N\,.
}
In section~\ref{sec_iib} we specialized our  discussion to four-dimensional 
$\mathcal N=1$ theories coming from compactifications of type IIB string theory
on general Calabi-Yau orientifolds with geometric and non-geometric fluxes. 
Our main result in this section is an explicit expression for the matrix $\mathsf Q$ mentioned 
above, however, we were not able to determine analytic expressions for its singular 
values for general flux configurations.


\subsubsection*{(Non-)integer conformal dimensions}

In section~\ref{sec_conf_dim} we therefore consider two particular cases for which 
we can determine the singular values $\sigma_{\alpha}$ analytically and numerically, respectively. 
Here we are interested in the question of what features of the compactification 
lead to the integer-valued conformal dimensions observed in 
\cite{Conlon:2021cjk,Apers:2022zjx,Apers:2022tfm,Quirant:2022fpn}.
\begin{itemize}

\item In section~\ref{sec_iia_mirror} we study the mirror-dual of 
the type IIA  DGKT construction \cite{DeWolfe:2005uu}. 
On the type IIB side we have good control over perturbative 
corrections in the large-complex-structure limit and were able 
to take them into account for the computation of the masses and 
conformal dimensions. 
As summarized in table~\ref{md_001}, we find that mass eigenvalues (in units of the AdS radius)
and conformal dimensions are in general not integer-valued --- only in the strict 
large-complex-structure limit we obtain integer conformal dimensions.

\item In section~\ref{sec_numerics} we consider a concrete type IIB example with 
one complex-structure modulus, one K\"ahler modulus, and the axio-dilaton.
We ignore corrections to the large-complex-structure limit and 
stabilize moduli using  geometric and non-geometric fluxes. 
For a choice of fluxes mirror-dual to the DGKT setting we find
integer conformal dimensions --- as expected --- however, when considering 
a slightly more general flux choice the masses (in units of $R^2_{\rm AdS}$) 
and conformal dimensions are no longer integer-valued.

\end{itemize}
Our findings therefore suggests that integer conformal dimensions occur a) for a specific 
choice of fluxes for which the superpotential splits into two separate terms, i.e.~the DGKT setting, and b) 
when ignoring perturbative 
corrections to the large-complex-structure limit. When deviating from either of those properties 
the conformal dimensions are in general no longer integer-valued. 
However, in our analysis we focussed only on the closed-string sector.
It would be interesting to take into account the open-string sector and repeat
the computation of masses and conformal dimensions.


\subsubsection*{Moduli stabilization in asymptotic regions}

Our discussion in  section~\ref{app_gen} is independent of our analysis 
of conformal dimensions, but uses many results from sections~\ref{sec_gen}
and \ref{sec_iib}.
We consider type IIB flux compactifications with only geometric $F_3$- and $H_3$-fluxes
in the large-volume limit. These configurations are relevant for the KKLT and large-volume scenarios. 
We compute the trace of the canonically-normalized mass matrix and argue 
that when stabilizing complex-structure and axio-dilaton moduli 
in an asymptotic region of moduli space at least one mass eigenvalue will diverge. 
This observations highlights that when stabilizing moduli one does not 
only need to ensure that the lightest modes are sufficiently heavy --- but also 
that the heaviest modes are separated from the Kaluza-Klein scale.
This point has recently been emphasized also in \cite{Blumenhagen:2022dbo}.


\clearpage
\subsection*{Acknowledgments}

We thank
Lucas Draijer,
Thomas Grimm,
Stefano Lanza,
Filippo Revello,
Maarten Rottier,
Lorenz Schlechter, and
Stefan Vandoren 
for very helpful discussions. 
The work of EP is supported by a Heisenberg grant of the
\textit{Deutsche Forschungsgemeinschaft} (DFG, German Research Foundation) 
with project-num\-ber 430285316.


\appendix 


\section{Computational details}
\label{app_conf}

In this appendix we summarize some details relevant for 
the computation of masses and conformal dimensions in 
section~\ref{sec_iia_mirror}.


\subsection{The computation of $\mathsf Q$}
\label{app_details}

Let us  explain the computation of the matrix $\mathsf Q$ shown 
in equation \eqref{res_q_001}.
We first note that the F-term conditions  $F_i=0$ for the superpotential 
\eqref{pot_w} take the form
\eq{
  \label{f_term_002}
  0 = \int_{\mathcal X}\chi_i \wedge \left( F_3- \Xi_A \op \mathsf T^A \right) .
}
Next, we recall that the K\"ahler potential $K_{\rm K}$ shown in \eqref{lu_005} only depends 
on the imaginary parts of the moduli $\mathsf T^A$ and therefore 
\eq{
  K_{AB\vphantom{\ov A}} 
  = \partial_{A\vphantom{\ov A}} \partial_{B\vphantom{\ov A}} K = - \partial_{\ov A}\partial_{B\vphantom{\ov A}} K
  = - G_{\ov A B}\,.
}
Using this relation and the fact that the superpotential \eqref{pot_w} is linear in $\mathsf T^A$ we obtain
\eq{
  \mathsf Q^{\ov A}{}_j = G^{\ov A C} \partial_C F_B = G^{\ov A C} \Bigl[ 
  K_{CB} - K_B K_C \Bigr] W
  = \Bigl[ - \delta^{\ov A}{}_B - K^{\ov A} K_B \Bigr]\op W \,.
}
With the help of the relation \eqref{rel_3846} we also determine
\eq{
  \mathsf Q^{\ov A}{}_j = G^{\ov A B} \partial_B F_j = G^{\ov A B} \int_{\mathcal X} \chi_j \wedge (-\Xi_B)
  = -i \op e^{-K_{\rm cs}} \op G^{\ov A B}  \, \Xi^{\ov i}_B\op G_{\ov i j}\,,
}
and along similar lines the expression for $\mathsf Q^{\ov i}{}_B$ is found. Finally, 
with $\ov\chi_{\ov i}$ the complex conjugate of $\chi_i$, 
we recall from 
\cite{Candelas:1990pi} that 
\eq{
  D_i \op \chi_j = -i\op e^{K_{\rm cs}} \op \kappa_{ij}{}^{\ov m} \ov\chi_{\ov m} \,,
}
where indices are raised by the inverse K\"ahler metric $G^{i\ov j}$. 
(This result was used recently also in \cite{Becker:2022hse} to determine 
the mass matrix.)
We then compute
using 
\eqref{lu_006}
and the complex-conjugates of \eqref{f_term_002} and \eqref{rel_3846}
\eq{
  \mathsf Q^{\ov i}{}_j = G^{\ov i k}\partial_k F_j  &= 
  G^{\ov i k} \int_{\mathcal X} D_k\op \chi_j \wedge \left( F_3- \Xi_A \op \mathsf T^A \right)
  \\
  &= -i \op e^{K_{\rm cs}}\op \kappa^{\ov i}{}_j{}^{\ov m} \int_{\mathcal X}\ov\chi_{\ov m} 
  \wedge \left( F_3- \Xi_A \op \mathsf T^A \right)
  \\
  &= -i \op e^{K_{\rm cs}}\op \kappa^{\ov i}{}_j{}^{\ov m} \int_{\mathcal X}\ov\chi_{\ov m} 
  \wedge \left( F_3- \Xi_A \op \ov{\mathsf T}^A 
  - \Xi_A \op \bigl(\mathsf T^A- \ov {\mathsf T}{}^A \bigr)
  \right)
  \\
  &= -i \op e^{K_{\rm cs}}\op \kappa^{\ov i}{}_j{}^{\ov m} \int_{\mathcal X}\ov\chi_{\ov m} 
  \wedge \Xi_A \op K^A
  \\
  & = \kappa^{\ov i}{}_{jn}\, \Xi^n_{\ov A} \op K^{\ov A}\,.
}


\subsection{Eigenvalues and eigenvectors of $\mathsf Q\ov{\mathsf Q}$}
\label{app_ev_details}

In this appendix we summarize the exact expressions for the eigenvalues and 
eigenvectors shown in table~\ref{ev_001}. 
In particular, the eigenvalues in the last two lines are given by
\eq{
\frac{\sigma^2_{(1)}}{\lvert W\rvert^2} & = 
\scalebox{0.674}{$
\frac{-5 \sqrt{\gamma  (\gamma  (\gamma  (25 \gamma +28)+258)+100)+169}+\gamma  \left(14 \sqrt{\gamma  (\gamma  (\gamma  (25 \gamma +28)+258)+100)+169}+\gamma  \left(\gamma  (13 \gamma +28)+\sqrt{\gamma  (\gamma  (\gamma  (25 \gamma +28)+258)+100)+169}+222\right)-20\right)+97}{2 (\gamma +1)^4} 
$}
\\
&= 16 +\frac{96}{13}\op\gamma+\mathcal O(\gamma^2)\,,
\\[10pt]
\frac{\sigma^2_{(2)}}{\lvert W\rvert^2} & = 
\scalebox{0.674}{$
\frac{5 \sqrt{\gamma  (\gamma  (\gamma  (25 \gamma +28)+258)+100)+169}+\gamma  \left(\gamma  \left(\gamma  (13 \gamma +28)-\sqrt{\gamma  (\gamma  (\gamma  (25 \gamma +28)+258)+100)+169}+222\right)-2 \left(7 \sqrt{\gamma  (\gamma  (\gamma  (25 \gamma +28)+258)+100)+169}+10\right)\right)+97}{2 (\gamma +1)^4}
$}
\\
&= 81 -\frac{5400}{13}\op\gamma+\mathcal O(\gamma^2)\,.
}
The corresponding eigenvectors shown in the last two lines of table~\ref{ev_001} are characterized by 
two parameters of the form
\eq{
 \eta_{(1)} &= \frac{\gamma  (2-5 \gamma )+\sqrt{\gamma  (\gamma  (\gamma  (25 \gamma +28)+258)+100)+169}-11}{6 (\gamma +1)}
 \\
 &= \frac{1}{3} + \frac{25}{39}\op\gamma+\mathcal O(\gamma^2)\,,
 \\[10pt]
 \eta_{(2)} &= \frac{\gamma  (2- 5 \gamma )-\sqrt{\gamma  (\gamma  (\gamma  (25 \gamma +28)+258)+100)+169}-11}{6 (\gamma +1)}
 \\
 &=-4 + \frac{48}{13}\op\gamma+\mathcal O(\gamma^2)\,.
}


\clearpage
\nocite{*}
\bibliography{references}

\providecommand{\href}[2]{#2}\begingroup\raggedright\begin{thebibliography}{10}

\bibitem{DeWolfe:2005uu}
O.~DeWolfe, A.~Giryavets, S.~Kachru, and W.~Taylor, ``{Type IIA moduli
  stabilization},'' {\em JHEP} {\bf 07} (2005) 066,
  \href{http://xxx.lanl.gov/abs/hep-th/0505160}{{\tt hep-th/0505160}}.

\bibitem{Conlon:2021cjk}
J.~P. Conlon, S.~Ning, and F.~Revello, ``{Exploring the holographic
  Swampland},'' {\em JHEP} {\bf 04} (2022) 117,
  \href{http://xxx.lanl.gov/abs/2110.06245}{{\tt 2110.06245}}.

\bibitem{Apers:2022zjx}
F.~Apers, M.~Montero, T.~Van~Riet, and T.~Wrase, ``{Comments on classical AdS
  flux vacua with scale separation},'' {\em JHEP} {\bf 05} (2022) 167,
  \href{http://xxx.lanl.gov/abs/2202.00682}{{\tt 2202.00682}}.

\bibitem{Apers:2022tfm}
F.~Apers, J.~P. Conlon, S.~Ning, and F.~Revello, ``{Integer conformal
  dimensions for type IIa flux vacua},'' {\em Phys. Rev. D} {\bf 105} (2022),
  no.~10 106029, \href{http://xxx.lanl.gov/abs/2202.09330}{{\tt 2202.09330}}.

\bibitem{Quirant:2022fpn}
J.~Quirant, ``{Noninteger conformal dimensions for type IIA flux vacua},'' {\em
  Phys. Rev. D} {\bf 106} (2022), no.~6 066017,
  \href{http://xxx.lanl.gov/abs/2204.00014}{{\tt 2204.00014}}.

\bibitem{Marchesano:2021ycx}
F.~Marchesano, D.~Prieto, and J.~Quirant, ``{BIonic membranes and AdS
  instabilities},'' {\em JHEP} {\bf 07} (2022) 118,
  \href{http://xxx.lanl.gov/abs/2110.11370}{{\tt 2110.11370}}.

\bibitem{Grimm:2004uq}
T.~W. Grimm and J.~Louis, ``{The Effective action of N = 1 Calabi-Yau
  orientifolds},'' {\em Nucl. Phys. B} {\bf 699} (2004) 387--426,
  \href{http://xxx.lanl.gov/abs/hep-th/0403067}{{\tt hep-th/0403067}}.

\bibitem{Kachru:2003aw}
S.~Kachru, R.~Kallosh, A.~D. Linde, and S.~P. Trivedi, ``{De Sitter vacua in
  string theory},'' {\em Phys. Rev. D} {\bf 68} (2003) 046005,
  \href{http://xxx.lanl.gov/abs/hep-th/0301240}{{\tt hep-th/0301240}}.

\bibitem{Balasubramanian:2005zx}
V.~Balasubramanian, P.~Berglund, J.~P. Conlon, and F.~Quevedo, ``{Systematics
  of moduli stabilisation in Calabi-Yau flux compactifications},'' {\em JHEP}
  {\bf 03} (2005) 007, \href{http://xxx.lanl.gov/abs/hep-th/0502058}{{\tt
  hep-th/0502058}}.

\bibitem{Breitenlohner:1982jf}
P.~Breitenlohner and D.~Z. Freedman, ``{Stability in Gauged Extended
  Supergravity},'' {\em Annals Phys.} {\bf 144} (1982) 249.

\bibitem{Kachru:2004jr}
S.~Kachru and A.-K. Kashani-Poor, ``{Moduli potentials in type IIa
  compactifications with RR and NS flux},'' {\em JHEP} {\bf 03} (2005) 066,
  \href{http://xxx.lanl.gov/abs/hep-th/0411279}{{\tt hep-th/0411279}}.

\bibitem{Plauschinn:2020ram}
E.~Plauschinn, ``{Moduli Stabilization with Non-Geometric Fluxes \textemdash{}
  Comments on Tadpole Contributions and de-Sitter Vacua},'' {\em Fortsch.
  Phys.} {\bf 69} (2021), no.~3 2100003,
  \href{http://xxx.lanl.gov/abs/2011.08227}{{\tt 2011.08227}}.

\bibitem{Becker:2002nn}
K.~Becker, M.~Becker, M.~Haack, and J.~Louis, ``{Supersymmetry breaking and
  alpha-prime corrections to flux induced potentials},'' {\em JHEP} {\bf 06}
  (2002) 060, \href{http://xxx.lanl.gov/abs/hep-th/0204254}{{\tt
  hep-th/0204254}}.

\bibitem{Shelton:2005cf}
J.~Shelton, W.~Taylor, and B.~Wecht, ``{Nongeometric flux compactifications},''
  {\em JHEP} {\bf 10} (2005) 085,
  \href{http://xxx.lanl.gov/abs/hep-th/0508133}{{\tt hep-th/0508133}}.

\bibitem{Villadoro:2006ia}
G.~Villadoro and F.~Zwirner, ``{D terms from D-branes, gauge invariance and
  moduli stabilization in flux compactifications},'' {\em JHEP} {\bf 03} (2006)
  087, \href{http://xxx.lanl.gov/abs/hep-th/0602120}{{\tt hep-th/0602120}}.

\bibitem{Shelton:2006fd}
J.~Shelton, W.~Taylor, and B.~Wecht, ``{Generalized Flux Vacua},'' {\em JHEP}
  {\bf 02} (2007) 095, \href{http://xxx.lanl.gov/abs/hep-th/0607015}{{\tt
  hep-th/0607015}}.

\bibitem{Dasgupta:1999ss}
K.~Dasgupta, G.~Rajesh, and S.~Sethi, ``{M theory, orientifolds and G -
  flux},'' {\em JHEP} {\bf 08} (1999) 023,
  \href{http://xxx.lanl.gov/abs/hep-th/9908088}{{\tt hep-th/9908088}}.

\bibitem{Kachru:2002sk}
S.~Kachru, M.~B. Schulz, P.~K. Tripathy, and S.~P. Trivedi, ``{New
  supersymmetric string compactifications},'' {\em JHEP} {\bf 03} (2003) 061,
  \href{http://xxx.lanl.gov/abs/hep-th/0211182}{{\tt hep-th/0211182}}.

\bibitem{Hull:2004in}
C.~M. Hull, ``{A Geometry for non-geometric string backgrounds},'' {\em JHEP}
  {\bf 10} (2005) 065, \href{http://xxx.lanl.gov/abs/hep-th/0406102}{{\tt
  hep-th/0406102}}.

\bibitem{Plauschinn:2018wbo}
E.~Plauschinn, ``{Non-geometric backgrounds in string theory},'' {\em Phys.
  Rept.} {\bf 798} (2019) 1--122,
  \href{http://xxx.lanl.gov/abs/1811.11203}{{\tt 1811.11203}}.

\bibitem{Grana:2006hr}
M.~Grana, J.~Louis, and D.~Waldram, ``{SU(3) x SU(3) compactification and
  mirror duals of magnetic fluxes},'' {\em JHEP} {\bf 04} (2007) 101,
  \href{http://xxx.lanl.gov/abs/hep-th/0612237}{{\tt hep-th/0612237}}.

\bibitem{Strominger:1990pd}
A.~Strominger, ``{SPECIAL GEOMETRY},'' {\em Commun. Math. Phys.} {\bf 133}
  (1990) 163--180.

\bibitem{Candelas:1990pi}
P.~Candelas and X.~de~la Ossa, ``{Moduli Space of {Calabi-Yau} Manifolds},''
  {\em Nucl. Phys. B} {\bf 355} (1991) 455--481.

\bibitem{Herraez:2018vae}
A.~Herraez, L.~E. Ibanez, F.~Marchesano, and G.~Zoccarato, ``{The Type IIA Flux
  Potential, 4-forms and Freed-Witten anomalies},'' {\em JHEP} {\bf 09} (2018)
  018, \href{http://xxx.lanl.gov/abs/1802.05771}{{\tt 1802.05771}}.

\bibitem{Marchesano:2019hfb}
F.~Marchesano and J.~Quirant, ``{A Landscape of AdS Flux Vacua},'' {\em JHEP}
  {\bf 12} (2019) 110, \href{http://xxx.lanl.gov/abs/1908.11386}{{\tt
  1908.11386}}.

\bibitem{Draijer:2022}
L.~Draijer, ``{Constraints on moduli masses in Type IIB orientifold
  compactifications},'' Master's thesis, Utrecht University, the Netherlands,
  2022.

\bibitem{Rottier:2022}
M.~Rottier, ``{Mass Hierarchies and Scaling Scenarios for Perturbatively Flat
  Flux Vacua},'' Master's thesis, Utrecht University, the Netherlands, 2022.

\bibitem{Douglas:2006zj}
M.~Douglas and Z.~Lu, ``{On the geometry of moduli space of polarized
  Calabi-Yau manifolds},'' \href{http://xxx.lanl.gov/abs/math/0603414}{{\tt
  math/0603414}}.

\bibitem{Conlon:2006gv}
J.~P. Conlon, ``{Moduli Stabilisation and Applications in IIB String Theory},''
  {\em Fortsch. Phys.} {\bf 55} (2007) 287--422,
  \href{http://xxx.lanl.gov/abs/hep-th/0611039}{{\tt hep-th/0611039}}.

\bibitem{Plauschinn:2021hkp}
E.~Plauschinn, ``{The tadpole conjecture at large complex-structure},'' {\em
  JHEP} {\bf 02} (2022) 206, \href{http://xxx.lanl.gov/abs/2109.00029}{{\tt
  2109.00029}}.

\bibitem{Grana:2022dfw}
M.~Gra\~na, T.~W. Grimm, D.~van~de Heisteeg, A.~Herraez, and E.~Plauschinn,
  ``{The tadpole conjecture in asymptotic limits},'' {\em JHEP} {\bf 08} (2022)
  237, \href{http://xxx.lanl.gov/abs/2204.05331}{{\tt 2204.05331}}.

\bibitem{Crino:2022zjk}
C.~Crin\`o, F.~Quevedo, A.~Schachner, and R.~Valandro, ``{A database of
  Calabi-Yau orientifolds and the size of D3-tadpoles},'' {\em JHEP} {\bf 08}
  (2022) 050, \href{http://xxx.lanl.gov/abs/2204.13115}{{\tt 2204.13115}}.

\bibitem{Blumenhagen:2022dbo}
R.~Blumenhagen, A.~Gligovic, and S.~Kaddachi, ``{Mass Hierarchies and Quantum
  Gravity Constraints in DKMM-refined KKLT},''
  \href{http://xxx.lanl.gov/abs/2206.08400}{{\tt 2206.08400}}.

\bibitem{Becker:2022hse}
K.~Becker, E.~Gonzalo, J.~Walcher, and T.~Wrase, ``{Fluxes, Vacua, and Tadpoles
  meet Landau-Ginzburg and Fermat},''
  \href{http://xxx.lanl.gov/abs/2210.03706}{{\tt 2210.03706}}.

\end{thebibliography}\endgroup
\bibliographystyle{utphys}


\end{document}